\documentclass[fleqn,11pt]{article}

\usepackage{amsfonts,amsmath,amssymb}
\usepackage{latexsym,color,cite}

\usepackage[utf8]{inputenc}
\usepackage{graphicx} %, psfrag}
\def\onecol{\onecolumn \mathindent 2em}

\def\noi{\noindent}

\newcommand{\Title}[1]{\noi {{\Large\bf #1}}\\[1ex]}

\newcommand{\Author}[2]{\noi{\bf #1}\\[2ex]\noi{\normalsize\it #2}\\}

\newcommand{\Abstract}[1]{\vskip 2mm \begin{center}
        \parbox{16.4cm}{\small\noi #1} \end{center}\medskip}
\newcommand{\foom}[1]{\protect\footnotemark[#1]}

\def\email#1#2{\footnotetext[#1]{e-mail: #2}\addtocounter{footnote}{1}}

\def\nq{\hspace*{-1em}}

\def\qq{\qquad}
\def\cm{\hspace*{1cm}}

\def\lal{&&\nq{}}
\def\eq{Eq.\,}

\def\beq{\begin{equation}}
\def\eeq{\end{equation}}
\def\bear{\begin{eqnarray}}
\def\bearr{\begin{eqnarray} \lal}
\def\ear{\end{eqnarray}}
\def\earn{\nonumber \end{eqnarray}}

\def\yyy{\\[5pt] \lal }

\def\diag{\mathop{\rm diag}\nolimits}

\def\const{{\rm const}}
\def\Half{\dfrac 12}

\def\eqn#1{\eq\eqref{#1}}
\def\rf{\eqref}
\def\mn{_{\mu\nu}}
\def\MN{^{\mu\nu}}
\def\mN{_\mu^\nu}

\def\M{{\mathbb M}}
\def\R{{\mathbb R}}

\def\cF{{\mathcal F}}
\def\cL{{\mathcal L}}
\def\cO{{\mathcal O}}
\def\cP{{\mathcal P}}
\def\sph{spherically symmetric}
\def\ssph{static, spherically symmetric}
\def\bh{black hole}
\def\bhs{black holes}
\def\wh{wormhole}
\def\whs{wormholes}
\def\asflat{asymptotically flat}
\def\cosm{cosmological}
\def\emag{electromagnetic}

\def\Scw{Schwarz\-schild}
\def\RN{Reiss\-ner-Nord\-str\"om}
\def\Areg{A_{\rm reg}}

\begin{document}
\onecol

\bigskip\bigskip

\Title{Regular black holes as an alternative to black bounce}

\Author{K.A. Bronnikov\foom 1}
	{Center of Gravitation and Fundamental Metrology, VNIIMS, Ozyornaya ulitsa 46, Moscow 119361, Russia\\
	 Peoples' Friendship University of Russia, ulitsa Miklukho-Maklaya 6, Moscow, 117198, Russia\\
	 National Research Nuclear University ``MEPhI,'' Kashirskoe shosse 31, Moscow 115409, Russia}

\Abstract
   {The so-called black bounce mechanism of singularity suppression, proposed by Simpson and Visser, consists 
   in replacing the spherical radius $r$ in the metric tensor with $\sqrt{r^2 + a^2}$, $a = \const >0$. 
   This removes a singularity at $r=0$ and its neighborhood from space-time, and there emerges a regular 
   minimum of  the spherical radius that can be a wormhole throat or a regular bounce (if located inside a 
   black hole).  Instead, it is proposed here to make $r=0$ a regular center by proper (Bardeen type) 
   replacements in the metric, preserving its form at large $r$. Such replacements are applied to a class of 
   metrics satisfying the condition $R^t_t = R^r_r$ for their Ricci tensor, in particular, to the \Scw, \RN, and 
   Einstein-Born-Infeld solutions. A simpler version of nonlinear electrodynamics (NED) is considered, for 
   which a  black hole solution is similar to the Einstein-Born-Infeld one but is simpler expressed analytically. 
   All new regular metrics can be presented as solutions to NED-Einstein equations with radial magnetic fields. 
 }

% ===========================================   
\email 1 {kb20@yandex.ru}  

% ============================= sec 1
\section{Introduction}
% =============================

  In general relativity (GR) and its classical extensions, the presence of singularities is quite a
  common though undesirable phenomenon, where a theory itself shows the boundaries of its 
  applicability. In general, the researchers probably do not believe in the existence of 
  singularities in Nature and hope that they must be somehow suppressed by effects of quantum 
  gravity. However, the numerous models and approaches in quantum gravity, 
  being translated to the language of classical physics, produce quite different results, see, 
  e.g., \cite{QG1, QG2, QG3, kelly20, achour20, kunst20, ash20a, bambi13}, and a 
  discussion in \cite{we20}. Thus, when applied to \bhs, some models predict \bh-white hole 
  transitions  \cite{QG2, QG3, kelly20, achour20}, others describe scenarios to a nonzero constant 
  value of the spherical radius at late times of the evolution \cite{kunst20}, there also emerge 
  configurations without any horizons \cite{bambi13}, etc. 
  Different ways of quantum-gravity regularization of \bh\ singularities are also discussed in the 
  recent papers \cite{gingrich24, koshelev24,bueno24}.  One can conclude that quantum gravity
  at its present stage of development is not yet ready to produce clear and unique predictions. 
   
  Therefore, it looks natural that the proposal made by Simpson and Visser (SV)  \cite{simp18} 
  to obtain a regular \ssph\ metric from a singular one by replacing the spherical radius $r$ 
  with the expression $r(u) = \sqrt{u^2 + a^2}$, thus removing a singularity at $r=0$, has caused
  a significant interest and was followed by a number of extensions and discussions.\footnote
		{Here, $u \in \R$ is a new radial coordinate instead of $r$. The notation $r$ is here kept 
		for the quantity of evident geometric meaning, the spherical radius, $r = \sqrt{-g_{\theta\theta}}$, 
		other radial coordinates are denoted by other letters to avoid confusion.}  
  This simple trick may be an easy way to simulate possible quantum gravity effects in the framework
  of classical gravity, leaving aside any details of quantization methods. Besides, it turned out 
  that new geometries that emerge in this way can have their own features of interest.

  This proposal, being applied to the \Scw\ solution, results in the globally regular metric \cite{simp18}
\beq                    \label{S-SV}
		ds^2 = \bigg(1 - \frac{2M}{\sqrt{u^2 + a^2}}\bigg)dt^2 
			- \bigg(1 - \frac{2M}{\sqrt{u^2 + a^2}}\bigg)^{-1} du^2 
			- (u^2 + a^2) (d\theta^2 + \sin^2\theta d\varphi^2).
\eeq  
  At small values of the regularization constant $a$ relative to the \Scw\ mass $M$ ($a < 2M$), 
  (its smallness looks most natural), the metric \rf{S-SV} describes a \bh\ with two horizons at 
  $u = \pm \sqrt{4M^2 -a^2}$. Larger values of $a$ lead to an extremal regular \bh\ with a 
  horizon at $u=0$ (if $a =2M$) and a \wh\ with a throat at $u=0$ at still larger $a$.
  In the \bh\ case $a < 2M$, the minimum of $r(u)$, observed at $u=0$, is located in a 
  Kantowski-Sachs anisotropic cosmological region between two horizons, where $u$ is a 
  temporal coordinate. Thus at $u=0$ happens a bounce in one of the two scale factors, $r(u)$, 
  of this cosmology, called a {\it black bounce} as suggested in \cite{simp18}. (The other scale factor 
  in this cosmology is $2M/\sqrt{u^2 + a^2}  - 1$, it takes a maximum value at the same instant 
  $u=0$.) With a discussion of other families of space-time metrics obtained with the same trick,
  all of them received the slang name of ``black-bounce space-times'' or simply ``black bounces.''
  One can also recall that black bounces in the precise meaning of this term appear as a common 
  feature in many solutions of GR and its extensions in the presence of phantom scalar fields. Such geometries
  have been named ``black universes,'' these are \bh\ space-times containing beyond their horizon 
  an expanding universe that becomes isotropic at late times, see, e.g.,
  \cite{bk-uni05, bk-uni06, bk-uni12, clem09, azreg11, trap18}.
  
  A black-bounce regularization of the Reissner-Nordstr\"{o}m solution of GR was constructed in
  \cite{franzin21}. The same approach was used to obtain a wide class of regular \bh\ and \wh\ space-times 
  in \cite{lobo20}. The diverse geometries found in this manner have attracted much attention, and 
  further studies involved their rotating counterparts \cite{mazza21, xu21, shaikh21}, quasinormal modes,
  gravitational wave echoes at possible black hole/\wh\ transitions and gravitational lensing parameters
  \cite{churilova19, yang21, guerrero21, tsukamoto21, islam21, cheng21, kb20, tsukamoto20, lima21,
  nascimento20, franzin22,vagnozzi22,pedrotti24,s_ghosh22,s_ghosh23}.
    
  A separate issue is to present SV-like space-times as possible solutions to the equations of GR with 
  different field sources. For \ssph\ space-times, such representations were obtained in \cite{rahul22} and 
  \cite{canate22}, and it was shown \cite{rahul22} that a large class of such space-times are obtainable as 
  solutions to the Einstein equations with a combined source consisting of a minimally coupled phantom scalar 
  field with a self-interaction potential, and an \emag\ field within nonlinear electrodynamics (NED), 
  whereas NED alone or a scalar field alone are unable to form a necessary source. A phantom field is 
  necessary for the existence of a minimum of the spherical radius $r$, while a NED source 
  is required for adjusting the total stress-energy tensor (SET) $T\mN$. The explicit forms of scalar and 
  NED sources of SV-regularized \Scw\ and \RN\ metrics were obtained in \cite{rahul22}, along with their
  global structure diagrams, including metrics with three and four horizons. A similar method was 
  applied to some cosmological space-times in \cite{kam22, kam24}.

  SV-like regularizations for other two families of singular solutions of GR were constructed in 
  \cite{kb22}: these were Fisher's solution with a massless canonical scalar field \cite{fisher48} and a 
  subset of dilatonic \bh\ solutions with interacting massless scalar and \emag\ fields 
  \cite{dil1, dil2, dil3, dil4}. In both cases, the SV substitution was applied in the simplest possible way 
  ($x \mapsto \sqrt{u^2 + a^2}$) to the factor $x$ that produced a space-time singularity at its zero 
  value. Scalar-NED sources for the regularized versions of these space-times were also found, and it turned
  out that that such a scalar field cannot not be only canonical (with positive kinetic energy) or only phantom 
  (with negative kinetic energy), but has to change its nature from one region to another, in other words,
  demonstrated what had been previously called  a ``trapped ghost'' behavior \cite{trap10, don11}. 
  The possible role of such fields in the stability properties of \bh\ and \wh\ space-times 
  was discussed in \cite{trap17, trap18}. More generally \cite{kb22}, a combination of NED and a minimally 
  coupled  scalar field (in general, of trapped ghost nature) in GR is able to provide a source for {\it any\/} 
  \ssph\ metric, while, according to \cite{kb23}, any such metric may be produced (though only piecewise) 
  with a nonminimally coupled scalar field as the only source. 

  A general feature of the SV proposal and its extensions is that the singularity in a geometry under study 
  is simply removed from space-time together with its neighborhood, being replaced by a throat or a black
  bounce. It leads to more complex geometries and causal structures, which may be considered, from 
  different viewpoints, both as an advantage and a shortcoming. There is, however, a natural alternative 
  to this approach: to try, instead of removing the singularity location $r=0$, to convert it to a regular center. 
  There are a great number of stellar and field models with regular centers, in particular, with NED sources
  (see, e.g., \cite{kb01-NED, kb18-NED} and references therein) and those whose origin is ascribed to 
  vacuum properties of various quantum fields including the gravitational one, see, e.g., 
  \cite{dym92,r_ghosh23} and references therein.
  However, our goal here is not to construct nonsingular models from the outset but to try to cure the 
  already existing singularities at $r=0$ by introducing small regularizing 
  parameters which may be hopefully ascribed to quantum gravity effects. As such examples, we will 
  consider metrics obeying the condition $R^t_t = R^r_r$ for their Ricci tensors since it is the property
  of many most important solutions of GR (which previously received regularization by the SV method): the 
  \Scw, \RN, their extensions with a \cosm\ constant and some others. Moreover, if we construct a regular 
  metric that preserves the property $R^t_t = R^r_r$, its source can be constructed with NED alone, 
  with no need for others kinds of matter.  
  
  The paper is organized as follows. In Section 2 we make some preparations, recalling the regular 
  center conditions and the way to obtain NED sources for the metrics under consideration.
  Section 3 is devoted to regular versions of the \Scw\ and \RN\ solutions and finding their pure NED sources. 
  In Section 4 we discuss a possible regularization of the Einstein-Born-Infeld space-time and one
  more solution of GR with a new NED resembling the Born-Infeld one but leading to simple analytical 
  expressions. Section 5 contains some concluding remarks. The metric signature $(+\,-\,-\,-)$ is adopted, 
  along with geometrized units such that $8\pi G = c = 1$.
   
% =================================== sec 2
\section{Preliminaries} 
% ===================================
\subsection{Regularity conditions}
% -----------------------------------------------------

  In this subsection we recall some well-known facts to be used in what follows.
  Consider a pseudo-Riemannian space-time $\M$ with an arbitrary \ssph\ metric 
\beq 		\label{ds}
		ds^2 = A(x) dt^2 - \frac{dx^2}{A(x)} - r^2(x) d\Omega^2,\qq 
		d\Omega^2 = d\theta^2 + \sin^2\theta d\varphi^2,
\eeq  
  written here in terms of the so-called quasiglobal radial coordinate $x$ \cite{BR-book}. This choice of
  the radial coordinate is well suited for the description of any \ssph\ space-times including \bhs\ 
  (where horizons appear as regular zeros of $A(x)$ provided $r(x)$ is finite) and \whs\ (where throats 
  appear as regular minima of $r(x)$ provided $A(x) > 0$). 
  
  If our space-time contains a location where $r\to 0$ under the condition $A > 0$, this location is called 
  a center, and it is indeed a center of symmetry in spatial sections of $\M$. If $r\to 0$ in a region where 
  $A < 0$ (as happens inside \bhs), it is a cosmological-type singularity instead of a center since $x$ can 
  there be used as a time coordinate. 
  
  A center is regular if all algebraic curvature invariants are there finite and smooth, which includes, 
  in particular, the existence of a tangent flat space-time at this point. For the metric \rf{ds} it
  implies that at some $x \to x_0$,
\beq             \label{reg-c}
			A(x) = A_0 + \cO(r^2), \qq  A(x) r'{}^2(x) = 1 + \cO(r^2), \qq  A_0 = \const >0,
\eeq   
  where the prime denotes $d/dx$, and the symbol $\cO(r^2)$ means a quantity of the same order as 
  $r^2$ or smaller. The second condition provides a correct circumference to radius ratio for small
  circles around the center.

% -------------------------------------------------------------------------------------------------
\subsection{Space-times with $R^t_t = R^x_x$ and their NED sources}
% -------------------------------------------------------------------------------------------------

 It makes sense to single out the important case of space-times where the Ricci tensor satisfies the 
 condition $R^t_t = R^x_x$. Then, by the Einstein equations
\beq 			\label{EE}
		G\mN \equiv R\mN - \Half \delta\mN R = - T\mN, 
\eeq  
  the SET of matter satisfies the same condition, and it holds not only for vacuum and a \cosm\ constant 
  but also for NED under spherical symmetry. Moreover, with the metric \rf{ds}, the equality 
  $R^t_t = R^x_x$ leads to the condition $r''(x) =0$, and almost without loss of generality we can 
  put $r(x) \equiv x$  (we thus only reject ``flux tubes'' with $r= \const$), so the quasiglobal coordinate
  $x$ coincides  with the more frequently used \Scw\ coordinate $r$, and we are dealing with the metric
\beq
		ds^2 = A(r) dt^2 - \frac{dr^2} {A(r)} - r^2 d\Omega^2,
\eeq  
  Since now $r' \equiv 1$, in the regularity conditions \rf{reg-c} we must put $A_0=1$.  
  
  Now, the only two nontrivial components of the Einstein equations read (the prime denotes $d/dr$)
\bearr                   \label{EE0}
  	 G^t_t = G^r_r = \frac{1}{r^2} [-1 + A + r A']  = - T^t_t,,
\yyy           			\label{EE2}
	 G^\theta_\theta = G^\varphi_\varphi = \frac{1}{2r} [r A'' + 2 r'A'] = - T^\theta_\theta,
\ear   

  This structure of the Ricci and Einstein tensors, hence the SET, can be represented by \sph\
  NED fields. In particular, if we consider the NED Lagrangian in the form $-\cL(\cF)$, where
  $\cF = F\mn F\MN$, and $F\mn$ is the \emag\ field tensor, the SET is in general  
\bearr                    \label{SET-F}
			T\mN[F] = - 2 \mathcal{L_F} F_{\mu\sigma} F^{\nu\sigma} 
					+\frac 12 \delta\mN \mathcal{L(F)},
\ear 
  with $\mathcal{L_F} = d\cL/d\cF$, and the \emag\ field equations are 
\beq		\label{eq-F}
			\nabla_\mu(\mathcal{L_F}F\MN) = 0.
\eeq
  With the present space-time symmetry, we may consider only radial electric and magnetic 
  fields, the only nonzero components of $F\mn$ being $F_{rt}=-F_{tr}$ and 
  $F_{\theta\varphi} =-F_{\varphi\theta}$. Let us suppose the existence of only the magnetic 
  components, such that
\beq
		F_{\theta\varphi} =-F_{\varphi\theta}= q \sin\theta, 
\eeq  
  where $q$ is a monopole magnetic charge. Then \eq \rf{eq-F} is trivially satisfied, while the invariant 
  $\cF$ is expressed as $\mathcal{F} = 2 q^2/r^4$, independently from the choice of $\cL(\cF)$. 
  The \emag\ SET takes the form
\bearr         \label{T-F}		
		T\mN[F] = \frac 12 \diag\Big(\cL,\ \cL,\ \cL - \frac{4q^2}{r^4} \cL_\cF,\
				\cL - \frac{4q^2}{r^4} \cL_\cF\Big). 
\ear  
  
  The way of obtaining the corresponding solutions with magnetic fields is described in a number of 
  papers devoted to GR-NED regular \bhs, e.g., \cite{kb01-NED, kb18-NED} and references therein.
  Thus, \eqn{EE0} may be presented in the integral form
\beq            \label{A(r)}
			A(r) = 1 - \frac{2M(r)}{r}, \qq  M(r) = \Half \int \rho(r) r^2 dr, 
\eeq   
  where $M(r)$ is called the mass function, $\rho (r) = T^t_t$ is the matter density, and thus 
  $\cL(\cF) = 2\rho$ can be found as a function of $r$ as follows:
\beq         \label{L(r)}
			\cL(\cF) =  \frac{2}{r^2} (1 - A - r A'),
\eeq  
  and it can be verified that the derivative $\cL_{\cF} = \cL'/\cF'$ calculated from \rf{L(r)} coincides 
  with $\cL_{\cF}$ determined using \eqn{EE2}, as must be the case. Moreover, in such solutions, as 
  should happen in regular magnetic \bhs\ \cite{kb01-NED}, $\cL(\cF)$ is finite at the center. 
  Let us remark that to obtain a solution regular at $r=0$, the integration in \rf{A(r)} must be carried out 
  from 0 to $r$, which leads to a total mass of purely \emag\ origin.
   
  One might consider NED sources with a radial electric field instead of a magnetic one,
  as is done, in particular, in \cite{we24} for a number of black-bounce space-times. This may be 
  implemented using, in addition to $\cF$, the auxiliary \emag\ invariant  $\cP = \cF \cL^2_\cF$, 
  and the whole construction does not look much more complicated than with magnetic 
  fields. It is, however, necessary to mention that electric NED solutions generically involve 
  different Lagrangians $\cL(\cF)$ in different regions of space, and this happens each time
  when $\cP(\cF)$, or conversely, $\cF(\cP)$ is not a monotonic function. As was observed in 
  \cite{we24}, in electric sources of black-bounce space-times this ambiguity does emerge, but 
  not in every solution. Unlike that, for space-times with a regular center we can predict that the 
  multivaluedness of $\cL(\cF)$ will necessarily emerge, in full analogy with the reasoning used in 
  \cite{kb01-NED, kb01-comm}. Indeed, in such cases $\cP(x) = -2q_e^2/r^4$ is a monotonic 
  function ($q_e$ is an electric charge), while $\cF(x)$ is not since it has to vanish both at  
  infinity and at a regular center.

% ========================================== sec 3
\section{Curing singularities in \Scw\ and \RN\ solutions}
% ==========================================
\subsection{General considerations}
% ----------------------------------------------

  The regularity conditions now reduce to the requirement $A(r) = 1 + \cO(r^2)$ at small $r$.
  The only function to be modified is $A(r)$ obeying \eqn{A(r)}, from which it follows that a regular 
  center corresponds to a finite value of $\rho(0)$. Let us assume that in a singular metric to be cured,
\beq            \label{A-sing}
			A(r) \approx 1 + A_1/r^m \qq {\rm as}\quad r\to 0, \qq 
				A_1 = \const,\quad m=\const  > 0.
\eeq.
  It is then easy to verfy that the substitution in the argument of $A(r)$
\beq            \label{r-reg}
			r \mapsto \frac{(r^2+a^2)^{n +1/2}}{r^{2n}}, \qq   
			n = \const \geq 1/m,   \qq a = \const > 0
\eeq
  leads to $A(r) \approx 1 + A_1 b^{-m (2n+1)} r^{2mn}$ at small $r$, with $2m n \geq 2$. A subtle 
  point is that in \rf{A-sing} the number $m$ is the {\it smallest\/} power in an expansion of $A$ in 
  powers of $1/r$, even though the singular asymptotic behavior of $A(r)$ is determined by the largest 
  power in this expansion.  
  
  It is easy to find that if $n = 1/m$, the resulting density $\rho(0)$ is nonzero, and the metric near 
  $r=0$ is asymptotically de Sitter if $\rho(0) > 0$ and AdS if $\rho(0) < 0$. If $n > 1/m$, then 
  $\rho(0) = 0$, and the metric near $r=0$ is asymptotically Minkowskian. One can notice that in
  the case $m = n = 1$ the substitution \rf{r-reg} actually coincides with the one used by Bardeen 
  in \cite{bardeen68} to convert a \Scw\ \bh\ to a regular one.The angular part of the metric does 
  not change, and with any value of $n$ the metric ``cured'' with \rf{r-reg} remains at large values 
  of $r$  approximately the same as the original, singular one. 
  
  We see that the recipe \rf{r-reg} smooths out singularities of the form \rf{A-sing} with $m > 0$, 
  producing different regular near-center density profiles depending on the constant $n$. 
  There can be, however, a ``softer'' singularity characterized by $m = 0$, it requires a somewhat 
  finer approach to be considered in the next section.
  
% --------------------------------------------------------------------------------------------------
\subsection{The \Scw\ metric} 
% --------------------------------------------------------------------------------------------------

  In the \Scw\ vacuum solution we have $A = 1- 2M/r$, where $M = \const$ is the mass parameter.
  In the above scheme it corresponds to $m =1$, and applying \rf{r-reg} with $n=1$, we obtain the 
  regularized function $A = \Areg (r)$ of the form suggested by Bardeen \cite{bardeen68} in his first
  regular \bh\ model,
\beq             \label{Scw1}
			\Areg (r) = 1 - \frac {2M r^2}{(r^2 + a^2)^{3/2}}.
\eeq  
  From \rf{L(r)} we then find the Lagrangian function of the corresponding NED source
\beq
			\cL(\cF)  = \frac{12 a^2 M}{(r^2 + a^2)^{5/2}} 
					    = \frac{12 a^2 M \cF^{5/4}}{(\sqrt{2q^2} + a^2 \sqrt{\cF})^{5/2}},
\eeq  
  coinciding with that obtained in \cite{ABG00} (up to notations). 

  With the function \rf{Scw1}, the metric is asymptotically de Sitter at $r=0$. An  asymptotically 
  Minkowskian metric at $r=0$ can be obtained with any $n > 1$ in \rf{r-reg}. For example, choosing 
  $n=3/2$, we find
\beq             \label{Scw2}
			\Areg(r) = 1 - \frac {2M r^3}{(r^2 + a^2)^2}, 
\qq % \nnn
			\cL(\cF)  = \frac{16 a^2 M r}{(r^2 + a^2)^3} 
					   = \frac{16 a^2 M (2q^2)^{1/4} \cF^{5/4}}{(\sqrt{2q^2} + a^2 \sqrt{\cF})^3},	
\eeq

  Note that the function $\cL(\cF)$ tends to a finite limit as $\cF \to \infty$ (to zero if $n>1$), as 
  should be the case at a regular center \cite{kb01-NED}, but does not have a correct Maxwell 
  limit ($\cL \sim \cF$) at small $\cF$.
  
  The behavior of $\Areg(r)$ is quite generic for regular NED-GR solutions, as can be seen in Fig.\,1: 
  it is a soliton-like structure at small masses, at some critical value of $M$ emerges an extremal 
  horizon, and at larger $M$ there is a \bh\ with two horizons and a global structure similar to
  the \RN\ one. Quite naturally, a plot of $\Areg(r)$ for an asymptotically Minkowski (at $r=0$) metric
  has a flattened summit, unlike the one for an asymptotically de Sitter metric, as illustrated in Fig.\,1c.  
%% ------------------------------------- fig 1
\begin{figure*} \centering   
\includegraphics[scale=0.3]{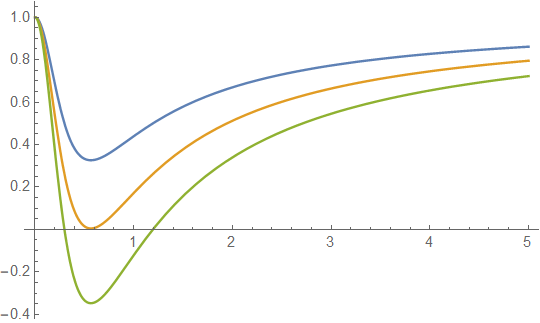}
\includegraphics[scale=0.3]{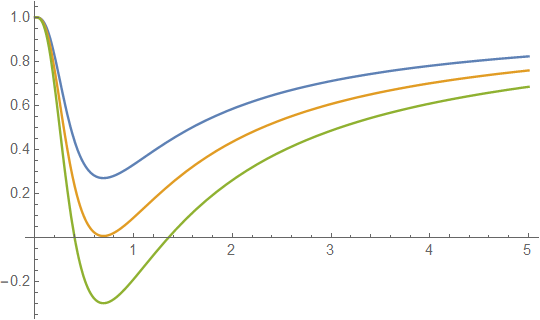}           
\includegraphics[scale=0.3]{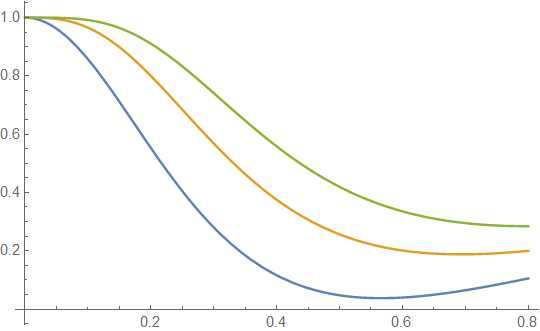}\\
		a \hspace{2in}  b \hspace{2in}  c           
\caption{\protect\small 
		Plots of $A(x)$ for the regularized \Scw\ metric with $a=0.4$: 
		(a) the function \rf{Scw1} with $M= 0.35, 0.517, 0.7$ (upside-down);
		(b) the function \rf{Scw2} with $M= 0.45, 0.612, 0.8$ (upside-down);
		(c) comparison of $\Areg(x)$ with the same $M = 0.5$ and $n = 1, 3/2, 2$;
		flattened plots at $r=0$ correspond to asymptotically Minkowski metrics..  
					}            \label{fig-A-Scw}            
\end{figure*}                                           
% ---------------------------------------        
  
% --------------------------------------------------------------------------------------------------
\subsection{The \RN\ metric} 
% --------------------------------------------------------------------------------------------------

  The \RN\ electrovacuum solution corresponds to the metric function 
\beq                \label{A-RN}
  	A = 1- 2M/r + Q^2/r^2, \qq M = {\rm mass}, \qq Q = {\rm charge}.
\eeq  	
  We have again $m=1$, and using the Bardeen replacement \rf{r-reg} with $n=1$, we obtain 
\beq        \label{Areg-RN}
		\Areg(r) = 1 - \frac {2M r^2}{(r^2 + a^2)^{3/2}} + \frac{Q^2 r^4}{(r^2 + a^2)^3},
\eeq
  with a de Sitter behavior near the regular center $r=0$. For the corresponding NED source
  we have according to \rf{L(r)},
\beq         \label{L-RN}
		\cL(r) = \frac {2 \big[6 a^6 M + 12 a^4 M r^2 + Q r^4 \sqrt{r^2 + a^2} 
						+ a^2 \big(6 M r^4 - 5 Q r^2 \sqrt{r^2 + a^2}\big)\big]}{(r^2 + a^2)^{9/2}};
\eeq  
  here and henceforth in similar formulas,an expression for $\cL(\cF)$ is readily obtained by substituting 
  $r^2 \mapsto \sqrt{2q^2/\cF}$ (please note that the charge $q$ refers to the NED source and 
  has nothing to do with the ``original'' charge $Q$ in the \RN\ metric). For convenience, to avoid 
  writing $|Q|$ in many relations, we assume $Q >0$.
  
  At small $r$ the function \rf{Areg-RN} behaves as 
\beq                \label{A0-RN}
		\Areg(r) = 1 - \frac{2M r^2}{a^3} + \frac{3a M +Q} {a^6}r^4 + \cO(r^6),
\eeq  
  so that the central asymptotic is de Sitter as long as $M > 0$. The case $M=0$ is treated 
  separately in Sec.\,3.5 since it requires another substitution for $A(r)$.  
  
  As a whole, the behavior of $\Areg(r)$ is rather diverse, as shown in Fig.\,2 for a particular value of 
  $a =0.4$, taken to be sufficiently large for illustration purposes. Two special cases deseving separate 
  attention are discussed below.   
%% ------------------------------------- fig 2
\begin{figure*} \centering   
\includegraphics[scale=0.3]{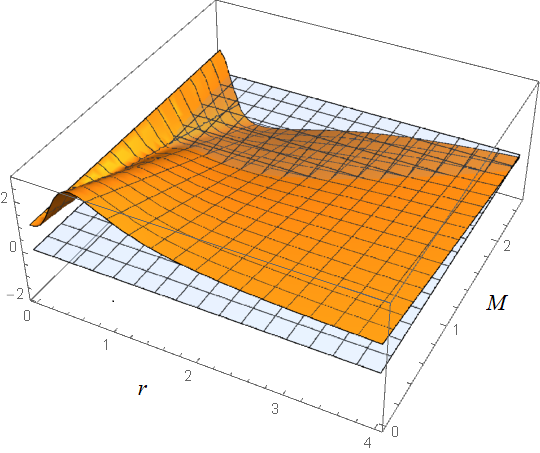}\cm
\includegraphics[scale=0.3]{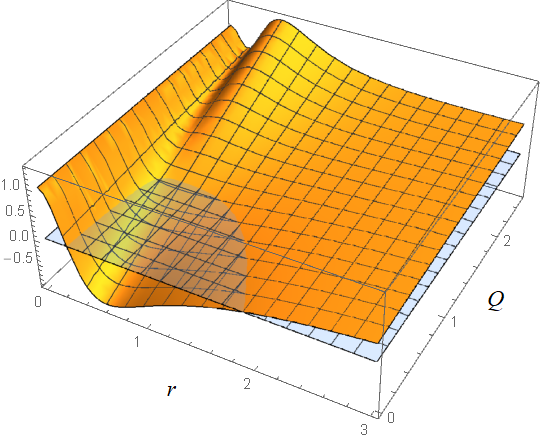}\\           
		a \hspace{2in}  b 
\caption{\protect\small 
		Plots of $\Areg(r)$ for the regularized \RN\ metric \rf{Areg-RN} with $a=0.4$: 
		(a) for fixed $Q =2$ and different $M$, (b) for fixed $M=1$ and different $Q$.
		The plane $A=0$ is shown in each panel to visualize the regions where $\Areg < 0$
		corresponding to \bh\ interiors. 
		}            \label{fig-A-RN}            
\end{figure*}                                           
% ---------------------------------------        

   Taking $n > 1$, we would obtain regularizations with a Minkowski central asymptotic, 
   with slightly other properties than \rf{Areg-RN} and \rf{L-RN}, not to be described ihere n detail.     
      
% --------------------------------------------------------------------------------------------------
\subsection{The extreme \RN\ metric} 
% --------------------------------------------------------------------------------------------------

  In the case $Q =M$, the function  \rf{A-RN} is a full square, $A(r) = (1 - M/r)^2$ (the same line 
  element also belongs to a \bh\ with a conformal scalar field \cite{bbm70, bek74, kb73}).
  As before, to obtain a regular metric with de Sitter behavior near $r=0$ we take $n=1$ and obtain 
\beq                  \label{Areg01}
		\Areg(r) = \bigg[1 - \frac {M r^2}{(r^2 + a^2)^{3/2}}\bigg]^2,
\quad \
		\cL(\cF)  =\frac{2 M \big[M r^4 + a^2 r^2 (-5 M + 6  \sqrt{a^2 + r^2})\big]
					+ 6 a^4 \sqrt{a^2 + r^2} }{(a^2 + r^2)^4}, 
\eeq
  while an asymptotically Minkowski center can be formed by choosing $n=3/2$, which leads to
\beq                  \label{Areg02}
		\Areg(r) = \bigg[1 - \frac {M r^3}{(r^2 + a^2)^2}\bigg]^2,
\qq
		\cL(\cF)  = \frac{2 M r (M r^5 + a^2 r^3 (-7 M + 8 r)+ 16 a^4 r^2 + 8 b^6 )}
					{(a^2 + r^2)^5}.
\eeq
   In both cases $\Areg(r)$ is nonnegative, but, depending on $M$ (at fixed $a$), it can have up to 
   two zeros corresponding to extremal horizons, as illustrated in Fig.\,3.
%% ------------------------------------- fig 3
\begin{figure*} \centering   
\includegraphics[scale=0.3]{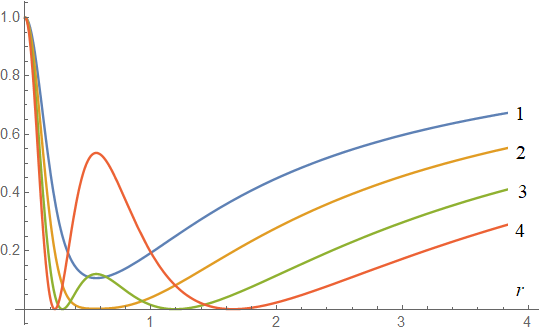}\qq
\includegraphics[scale=0.3]{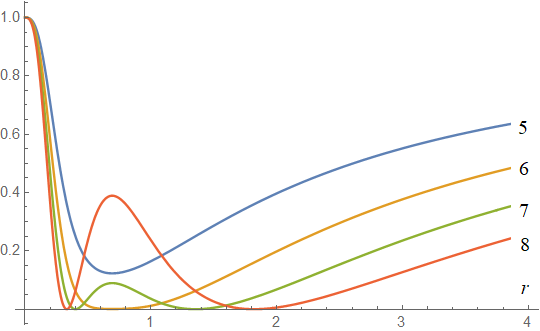}\qq
\includegraphics[scale=0.6]{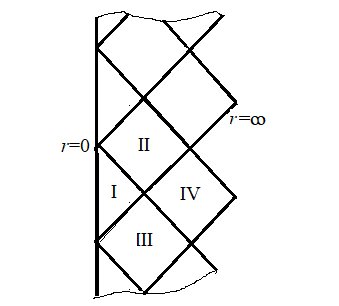}\\           
		a \hspace{2in}  b \hspace{2in}  c
\caption{\protect\small 
		Plots of $\Areg(r)$ for the regularized extremal \RN\ metric \rf{Areg-RN} with $a=0.4$: 
		(a) for the function \rf{Areg01} with $M = 0.7, 1, 1.4, 1.8$, curves 1-4 ;
		(b) for the function \rf{Areg02} with $M = 0.8, 1.2, i.6, 2$, curves 5-8. 
		Panel (c) shows the Carter-Penrose diagram for a configuration with two extremal horizons. 
		}            \label{fig-A-RNe}            
\end{figure*}                                           
% ---------------------------------------         
  The Carter-Penrose global structure diagram in Fig.\,3c for a space-time with two extremal 
  horizons indefinitely extends up and down; region I and its analogs are those near the center,
  region II and III are those between the horizons, and regions like IV are the external \asflat\
  ones.
   
% --------------------------------------------------------------------------------------------------
\subsection{The massless \RN\ metric} 
% --------------------------------------------------------------------------------------------------

  Consider the only case of \RN\ metric corresponding to \rf{A-sing} with $m = 2$, with
\beq                               \label{0-RN}
		A(r) = 1 + Q^2/r^2, \qq Q = \const = \text{charge}.
\eeq   
  Thus everywhere $A > 1$, and there is a naked repulsive singularity at $r=0$ 
  
  In \rf{r-reg}, to obtain an (A)dS behavior near $r=0$, we must take $n=1/2$, with the results 
\beq               \label{RN01}
		\Areg(r) = 1 + \frac {Q^2 r^2}{(r^2 + a^2)^2},
\qq
		\cL(\cF)  = \frac{2Q^2 (r^2 - 3a^2)}{(r^2 + a^2)^3}.
\eeq
	Near $r=0$ we have  $\cL(\cF)  = 2\rho < 0$, corresponding to an anti-de Sitter center,
	as illustrated in Fig.\,4a.  A Minkowski asymptotic behavior near $r=0$ is obtained, for example, 
	with $n=1$, which leads to
\beq               \label{RN02}
		\Areg(r) = 1 - \frac {Q^2 r^4}{(r^2 + a^2)^3},
\qq
		\cL(\cF)  = \frac{2Q^2 r^2 (r^2 - 5a^2)}{(r^2 + a^2)^4},
\eeq
   and again $\cL(\cF)  = 2\rho < 0$ near $r=0$, but it is zero at the center itself due to the factor 
   $r^2$  (Fig.\,4b).  It is the case where $\Areg(r)$ has the shape of a pure barrier, while 
   barriers along with depressions are observed in the more general pictures in Fig.\,2.
%% ------------------------------------- fig 4
\begin{figure*} \centering   
\includegraphics[scale=0.27]{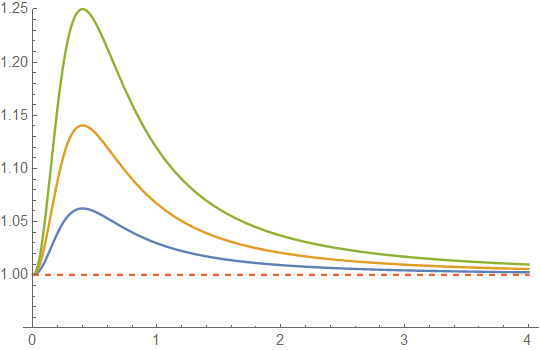}
\includegraphics[scale=0.27]{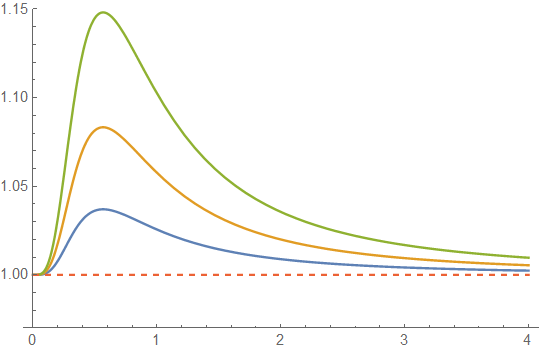}
\includegraphics[scale=0.27]{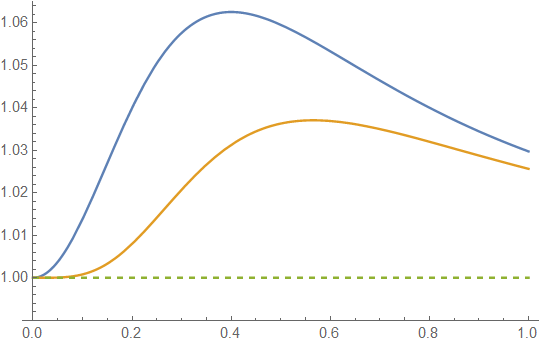}\\           
		a \hspace{2in}  b \hspace{2in} c
\caption{\protect\small 
		Plots of $\Areg(r)$ for the regularized massless \RN\ metric with $a=0.4$ and 
		$Q = 0.2,\ 0.3,\ 0.4$ (bottom-up in each panel): 
		(a) \eqn{RN01} with an AdS central asymptotic,
		(b) \eqn{RN02} with a Minkowski central asymptotic;
		(c) a comparison of two curves for $Q=0.2$ at small $r$, the flattened curve near 
		$r=0$ corresponding	to a Minkowski asymptotic.
		}            \label{fig-A-RN}            
\end{figure*}                                           
% ---------------------------------------     
    
% ========================================== sec 3
\section{Curing singularities in some Einstein-NED solutions}
% ==========================================    
    
  The metric \rf{A-sing} with $m =0$, with finite $A(0) \ne 1$, also has a singularity at $r=0$, such that
  $\rho \sim 1/r^2$ at small $r$, and the integral in \rf{A(r)} taken from zero to small $r$ behaves 
  as $\const \cdot r$ and adds a constant to $A$ according to \rf{A(r)}. In this case, replacements like 
  \rf{r-reg} do not work because there is no particular $r$ dependence of $A$ to be modified. 
  Instead, one can either directly modify $\cL(\cF)$ to lead it to a finite limit at large $\cF$ or make 
  finite the density $\rho(r)$, for example, replacing there $1/r^2$ with $1/(r^2 + a^2)$. Let us 
  consider two such examples.
 
% --------------------------------------------------------------------------------------------------
\subsection{The Einstein-Born-Infeld solution} 
% --------------------------------------------------------------------------------------------------

  The famous Born-Infeld NED Lagrangian, which for pure electric or magnetic fields reads
\beq              \label{L-BI}
		\cL(\cF) = \beta \Big(-1 + \sqrt{1 + 2\cF /\beta^2}\Big),\qq \beta = \const >0,
\eeq  
  is known to make finite the \emag\ field strength and energy in its Coulomb-like \sph\ solution 
  while, being coupled to GR, it weakens but does not remove the curvature singularity at $r=0$. Thus, 
  with \rf{L-BI}, integration in \rf{A(r)} with substituted $\cF = 2Q^2/r^4$ (assuming a radial 
  magnetic field with charge $Q$, leads to the expression (see, e.g., 
  \cite{breton02,fern05, kru17,fabris21})
\beq        \label{A-BI}
		A(r) = 1 - \frac 16 \Big(\sqrt{4Q^2 + \beta^2 r^4} - \beta r^2\Big) 
		-\frac{2}{3r}\sqrt{\frac{|Q|^3}{\beta}}(1+i)
		 F \bigg(i \sinh^{-1}\Big[(1+i)\frac{r\sqrt{\beta}}{2\sqrt{|Q|}}\Big], -1 \bigg),
\eeq  
  where $F$ is an elliptic function. The asymptotic behavior of $A(r)$ at small $r$ is 
\beq       \label{A-BI0}
		A(r) = 1 - \beta |Q| + \beta^2 r^2/6 + \cO(r^4)
\eeq  
  and can be associated with \rf{A-sing} at $m =0$. Examples of the behavior of $A(r)$ with $\beta =2$
  (used for convenience), both corresponding to a black hole (if $|Q| \geq 0.6$) and a naked singularity
  (otherwise), are shown in Fig.\,5a.  
  
  As discussed above, to regularize the metric in this case, it is necessary to modify the Lagrangian
  function for having its finite limit as $\cF \to \infty$. Such a simple way, preserving the Born-Infeld behavior    
  \rf{L-BI} of $\cL(\cF)$ at moderate values of $\cF$, can be proposed as 
\beq               \label{L-MBI}
		\cL(\cF) = \beta \bigg(-1 + \sqrt{1 + \frac {2\cF}{\beta^2+ \gamma\cF}}\bigg),
				\qq \beta, \gamma = \const >0,
\eeq   
  assuming sufficiently small values of $\gamma$. For $A(r)$ we then obtain a long expression  
  with several Appel functions of six arguments, not to be presented here, with the near-center behavior
\beq
		\Areg(r) = 1 - \frac 16 \Big(- \beta + \sqrt{\beta^2 + 2/\gamma}\Big)r^2 + \cO(r^4), 	
\eeq  
  indicating a de Sitter asymptotic. The plots of $\Areg(r)$, drawn in Fig.\,5b for particular values of the 
  parameters, show the characteristic form of the metric functions known in regular NED-GR \bh\ solutions.
%% ------------------------------------- fig 5
\begin{figure*} \centering   
\includegraphics[scale=0.27]{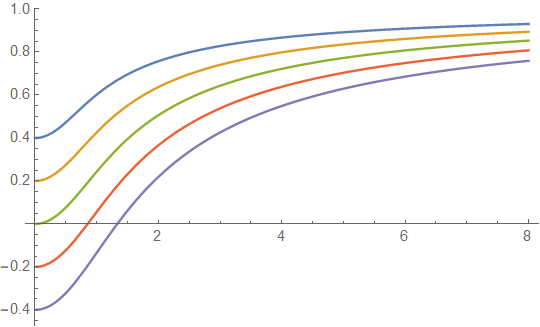}\cm
\includegraphics[scale=0.27]{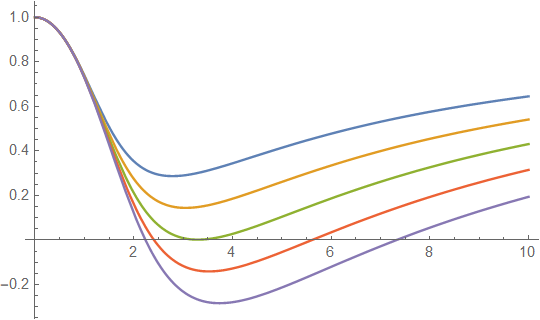}\\           
		a \hspace{2in}  b 
\caption{\protect\small 
		Plots of $A(r)$ in solutions with the original (a) and modified (b) Born-Infeld theory as a source 
		of gravity. The parameters are: (a) $\beta =2$, $q = 0.6, 0.8, 1, 1.2, 1.4$ (upside-down);
		(b) $\beta =2, \gamma =0.5, q= 2.5, 3, 3.5, 4, 4.5$ (upside-down). 
		}            \label{fig-A-BI}            
\end{figure*}                                           
% ---------------------------------------          
  
  Let us, for methodological purposes, introduce a NED theory, whose solutions behave similarly to 
  those of the Born-Infeld theory but are expressed analytically in a much simpler way.
  
% -----------------------------------------------------------------------------------
\subsection{A simpler NED with Born-Infeld-like behavior} 
% -----------------------------------------------------------------------------------
  
  Consider the NED Lagrangian (from a family of theories considered in \cite{kru17b})
\beq          \label{L-N1}
		\cL(\cF) = \frac{\cF}{1 + h\sqrt{\cF/2}}, \qq h = \const > 0.
\eeq  
  Like the Born-Infeld theory, this one takes a correct Maxwell form, $\cL = \cF$, at small $\cF$. 
  With $\cF = 2Q^2/r^4$, \eq\rf{A(r)} leads to the following expression for $A(r)$:
\beq          \label{A-N1}
		A(r) = 1 - \frac{Q^2}{Hr} \arctan \frac r H, \qq   H := (Q^2 h^2)^{1/4},
\eeq   
  with the near-center behavior similar to \rf{A-BI0},
\beq            \label{A-N10}
		A(r) = 1 - \frac{|Q|}{h} + \frac{r^2}{3 h^2} + \cO(r^4), 
\eeq  
  and the $r$ dependence of the Lagrangian function (equal to $2\rho$) is
\beq            
		\cL(\cF(R)) = \frac{2Q^2}{r^2 (r^2 + H^2)}.	
\eeq		  

  It is now easy to regularize the system by simply replacing $1/r^2 \mapsto 1/(r^2+c^2)$ in this $\cL(r)$,
  $c >0$ being a small regularization parameter:
\beq          \label{L-N1reg}
		  \cL(\cF) \ \mapsto \ \cL_{\rm reg} = \frac{2Q^2}{(r^2 + c^2) (r^2 + H^2)} 
		   		= \frac{\cF}{(1 + h\sqrt{\cF/2})\big(1 + c^2 |Q|^{-1/2}\sqrt{\cF/2}\big)}.
\eeq  
  This leads to the regularized metric function
\beq          \label{A-N1reg}
		\Areg(r) = 1 - \frac{Q^2}{(H^2 - c^2)r}\bigg(H \arctan \frac rH - c \arctan \frac rc \bigg),
\eeq
  with the near-center asymptotic behavior
\beq
		\Areg(r) = 1 - \frac{|Q| r^2}{3 c^2 h^{3/2}} + \cO(r^4).
\eeq  
  It can be verified that the functions \rf{A-N1} and \rf{A-N1reg} behave qualitatively in the same way
  as solutions for the Born-Infeld theory and its modification \rf{L-MBI}, shown in Fig.\,5.     
  
   One can notice that in both theories \rf{L-BI} and \rf{L-N1} the solution has either a naked singularity
   or a single horizon, whereas the regularized solution is either soliton-like or represents a regular \bh\ 
   with one or two horizons (a single, extremal horizon emerges in the intermediate case). This picture is
   common to many solutions with a regular center and is related to the fact that the regularity requires
   $A(0) =1$, hence such a center can only exist in a static region with $A>0$.    
       
% ======================== sec 5
\section{Concluding remarks}
% ========================

  Let us enumerate and discuss some results and observations made in this paper.
  
\begin{itemize}
\item  
  A way of singularity removal is proposed for \ssph\ space-times satisfying the condition $R^t_t = R^r_r$
  by properly changing a neighborhood of a singularity at $r=0$ so that it becomes a regular center.
  This is achieved by a Bardeen-like replacement in the metric function $A(r)$ containing a small 
  regularization parameter; the condition $R^t_t = R^r_r$ is preserved, and the resulting metric can 
  be interpreted as an Einstein-NED solution with a radial magnetic field, by analogy with studies devoted 
  to regular NED-sourced \bhs, see, e.g., \cite{kb01-NED,kb22-NED} and references therein.
\item   
  More involved are cases where the original singularity is of comparatively soft nature (for instance, 
  those with a finite value of $A(0) \ne 1$), its removal requires a direct modification in the original 
  Lagrangian or the expression for the energy density, as is seen in the example of the 
  Einstein-Born-Infeld solution. A much simpler version of NED is considered, given by \eqn{L-N1}, 
  which leads to similar properties of the solution near the singularity, while its modified version 
  \rf{L-N1reg} leads to a regular solitonic or \bh\ solution.  
\item
  We have dealt here with examples of \asflat\ space-times, with \RN\ behavior at large $r$.
  There is a straightforward extension of these solutions with a nonzero \cosm\ constant $\Lambda$, 
  similar to that described, e.g., in \cite{fern06}. It is achieved by adding the term $- \Lambda r^2/3$ 
  to all functions $A(r)$ mentioned in this paper, and this term even does not need any modification 
  at singularity removal since it does not spoil the metric behavior near a center. 
\end{itemize}  

  Black-bounce space-times contain regular minima of the spherical radius $r$ and therefore require 
  phantom matter as a source in the framework of GR. Obtaining regular centers instead of 
  singularities does not require such exotic matter, and, in particular, all metrics with $R^t_t = R^r_r$ 
  can be sourced by nonlinear \emag\ fields which (though marginally) satisfy the Null Energy Condition.
  
  More general space-times, not restricted by the above condition for the Ricci tensor, require 
  other regularization methods, and their formulation can be a next task. Concerning their possible 
  material sources, one may recall that, as shown in \cite{kb22}, any \ssph\ metric may be described 
  as a solution of GR with a combined source consisting of a magnetic field obeying NED and a scalar field 
  with a certain self-interaction potential, and one may hope that, unlike black-bounce space-times, those
  with a regular center will not require a phantom field as a source. 
  
   Various aspects of the new proposed regular metrics can be further studied, such as geodesics, lensing,
   stability, quasinormal modes, etc. In the context of finding sources of gravity for such metrics, 
   one can note that the stability of a particular geometry essentially depends on the dynamics of its 
   source. Thus, it has been found that the Ellis simple wormhole geometry \cite{ellis73, kb73} 
   can be stable or unstable, depending on the source it is ascribed to: a phantom scalar field, 
   a perfect fluid or a k-essence field \cite{gonz08, kb-sha13, kb-fa21}. 
   
\subsection*{Acknowledgment}
   The work supported by the Ministry of Science and Higher Education of the Russian Federation, Project 
   FSWU-2023-0073. Many thanks to my colleagues Milena Skvortsova and Sergey Bolokhov for numerous 
   helpful discussions.  
  
% ===========================

\end{document}